\documentclass[runningheads]{llncs}
\usepackage[T1]{fontenc}
\usepackage{comment}
\usepackage{multirow} 
\usepackage{marvosym} 
\usepackage[colorlinks, linkcolor=blue, citecolor=blue, urlcolor=blue]{hyperref}
\usepackage{graphicx}
\usepackage{booktabs}
\begin{document}
\title{Isotropic Remeshing with Inter-Angle Optimization\thanks{Supported by the National Key R\&D Program of China (2024YFB3908500, 2024YFB3908503), Guangdong Basic and Applied Basic Research Foundation (2023A1515110292), Scientific Foundation for Youth Scholars of Shenzhen. \textsuperscript{(\Letter)}Corresponding author.}}
%
%

\author{Hanbing Zheng \and
Chenlei Lv\textsuperscript{(\Letter)}}
\authorrunning{H. Zheng and C. Lv.}
%
\institute{College of Computer Science and Software Engineering, \\Shenzhen University, Shenzhen, China\\
\email{chenleilv@mail.bnu.edu.cn}}
\maketitle              

\begin{abstract}
\vspace{-4mm}
As an important metric for mesh quality evaluation, the isotropy property holds significant value for applications such as texture UV-mapping, physical simulation, and discrete geometric analysis. Classical isotropy remeshing methods adjust vertices and edge lengths, which exhibit certain limitations in terms of input data sensitivity, geometric consistency control, and convergence speed. In this paper, we propose an improved isotropy remeshing solution with inter-angle optimization during mesh editing to enhance shape control capability and accelerate convergence. The advantage of the solution lies in its ability to predict the impact of edge length adjustments on subsequent optimization by monitoring angle transformations. It avoids inefficient editing that may cause performance fluctuations, thereby improving efficiency. Experiments demonstrate that the proposed method effectively improves the overall efficiency of mesh optimization. (The code has been released at \href{https://github.com/vvvwo/Isotropic-Remeshing-InterAngle}{Isotropic-Remeshing-InterAngle})

\end{abstract}

\section{Introduction}

As a classical 3D data representation, triangular mesh models have been widely employed in various applications, including facial recognition~\cite{lv2019nasal,lv20203d}, digital infrastructure~\cite{zhang2024architectural,he2024windpoly}, gaming\&film production, and geographic information systems. Due to the excellent computability, adaptability to physical properties, and clear data structure, mainstream 3D geometry engines universally adopt triangular meshes as the standard data format for storage and computation. The initial triangular mesh is constructed either through shape reconstruction methods~\cite{lv2021voxel,kazhdan2013screened} or manual editing in CAD software. From the perspective of discrete differential geometry, shapes of triangular facets within the mesh exhibit randomness and often fail to satisfy the Delaunay condition~\cite{yi2018delaunay}, which is unfavorable for subsequent numerical computations. Therefore, isotropic remeshing methods are proposed for raw mesh optimization.

The concept of isotropic remeshing was proposed as early as the 1990s~\cite{bern1994provably}. The primary objective is to transform an unconstrained triangular mesh (anisotropic) into a new one composed of approximately equilateral triangles (isotropic) while preserving geometric consistency. There are two technical solutions for the target, including Centroidal Voronoi Tessellation (CVT)~\cite{levy2010p} and four-steps strategy~\cite{botsch2004remeshing}. CVT iteratively adjusts vertex positions to uniform vertex-based Voronoi cells by employing Lloyd's relaxation. Limitations include sensitivity to initial mesh, slow convergence rates, and high computational coupling. Four-steps strategy is to reconnect edges between vertexes with position adjustment. It contains four basic operations (split, collapse, flip, and tangent smoothing) to efficiently adjust vertex-based distances. Intuitively, edge reconnection appears more efficient, but the original scheme still exhibits deficiencies in result stability and convergence efficiency due to insufficient consideration of shape affection.

In this paper, we propose a new isotropic remeshing method based on four-steps strategy. We redesign shape constraints with inter-angle optimization, which ensures better shape controlling in edge reconnection and prevents inefficient editing operations. The inter-angle optimization enhances the efficiency of edge-based operations during each iteration while mitigating cross-interference between different edge operations. To better preserve geometric consistency, we additionally employ point cloud up-sampling with Moving Least Squares (MLS)~\cite{alexa2001point} remapping, ensuring adjusted vertices remain constrained to the original mesh-defined surface. Contributions can be summarized as:

\begin{itemize}

\item We present an inter-angle optimization scheme to control basic operations in the isotropic remeshing. It is useful to enhance convergence efficiency while mitigating aggressive shape deformation, thereby maintaining superior geometric consistency.

\item We propose a MLS-based solution for geometric consistency keeping. The proposed solution eliminates the need for tangent smoothing, preventing vertex optimization from deviating from the MLS surface. Particularly for sparse meshes, it guarantees the stability for the vertex adjustment.

\end{itemize}

\section{Related Works}

As mentioned before, there are two technical solutions for isotropic remeshing, including CVT-based and four-steps remeshing.

\textbf{\textit{CVT-based Remeshing.}} Such solutions employ the Voronoi Diagram or Delaunay triangulation to implement remeshing. Representative methods include Restricted Voronoi Diagram (RVD)~\cite{yan2009isotropic}, improved RVD~\cite{yan2014blue}, weighted diagram optimization~\cite{goes2014weighted}, intrinsic Delaunay remeshing~\cite{liu2015efficient}, geodesic-based CVT~\cite{ye2019geodesic}, parallel CVT~\cite{zheng2020computing}, and RVD with signed distance field (SDF)~\cite{hou2022sdf}. Such methods enhance the original CVT scheme by incorporating intrinsic metrics to suppress geometric artifacts induced by tangent-space displacements, thereby improving the stability of vertex neighborhood structures. However, such improvements inevitably introduce computational overhead, which compromises overall efficiency. Parallel optimization presents a potential solution, yet the inherently high coupling within CVT inevitably compromises isotropic quality.

\textbf{\textit{Four-steps Remeshing.}} Edge reconnection-based or four-steps remeshing implement mesh optimization based on four basic operations, including split, collapse, flip, and tangent smoothing. Such operations directly modify edge lengths and vertex degrees, which are more efficient than CVT-based remeshing. Representative methods include principal direction field-based remeshing~\cite{alliez2003anisotropic}, dynamic surface remeshing~\cite{jiao2010anisotropic}, adaptively isotropic remeshing~\cite{dunyach2013adaptive,lv2022adaptively}, implicit domain-based remeshing~\cite{dapogny2014three}, manifold-constrained remeshing~\cite{hu2021manifold}, and intrinsic\&isotropic remeshing~\cite{lv2022intrinsic}. In practice, some researchers have found that introducing shape control mechanisms into the four-steps strategy is essential for performance improvement. Wang {\em et al.}~\cite{wang2018isotropic} proposed an isotropic remeshing method by removing large and small angles. Xu {\em et al.}~\cite{xu2019anisotropic} designed a new edge reconnection strategy to reduce the occurrence of obtuse angles. It does not pursue the strictly isotropic property. Lv {\em et al.}~\cite{lv2022adaptively} employs angle-based constraints to judge the implementation of some edge editing operations. Such studies have established preliminary strategies for inter-angle optimization, which inspire our more comprehensive solution.

\section{Background}

As mentioned in early four-steps scheme~\cite{botsch2004remeshing}, basic operations for isotropic remeshing include split, collapse, flip, and tangent smoothing, which uniform edge length. Defining a regular edge length $l$, the split inserts a new vertex for a long edge ($>4/3l$) and the collapse merges a short edge ($<4/5l$). The flip remove a edge for two vertexes and add a new edge between their common neighbors, which optimizes the vertex-based degree. The tangent smoothing is to adjust vertex positions to achieve accurate isotropic property. 

Iteratively processing the four operations can implement the isotropic remeshing for a raw mesh. However, in the original implementation, coupling effects between the four operations are not taken into consideration. For instance, when performing collapse operations, the change in vertex degrees may increase the computational load for flips. Therefore, we aim to achieve more efficient control through inter-angle optimization at each step, thereby reducing mutual interference and enhancing the algorithm's convergence performance.

\section{Methodology}

We re-design basic operations (split, collapse, and flip) with more reasonable inter-angle optimization for triangular faces. The primary objective is to introduce more precise shape controlling to constrain remeshing operations and prevent "fixing one issue while creating another". In addition, we employ a MLS-based up-sampling to control vertex positions in the tangent smoothing for geometric consistency keeping. It is useful for some regions with few vertexes or sharp curvature changing.

\subsection{Inter-Angle Optimization}

The proposed inter-angle optimization provides shape constraints during the isotropic remeshing. According to different edge editing operations, we develop related inter-angle checking to predict potential shape degradation, thereby preventing the execution of detrimental mesh editing operations. Basically, the optimization includes three parts: split, collapse, and flip schemes.

\begin{figure}
\centering
\vspace{-2mm}
\includegraphics[width=0.65\textwidth]{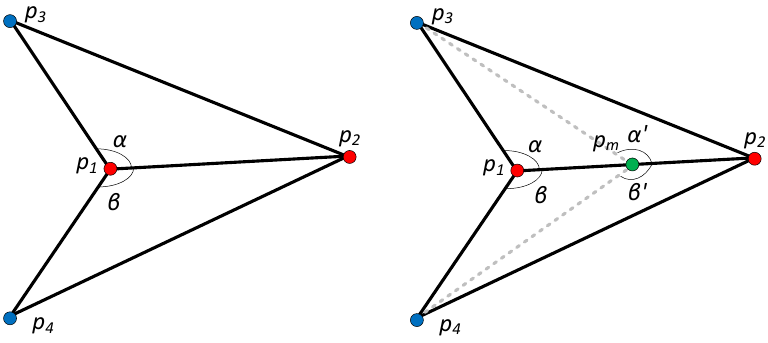}
\vspace{-2mm}
\caption{Inter-angle optimization for split scheme. Angles $\angle\alpha$ and $\angle\beta$ control the split operation for $\overline{p_1p_2}$, which prevent the formation of $\angle\alpha‘$ and $\angle\beta’$.} 
\vspace{-4mm}
\label{F1}
\end{figure}

\textbf{\textit{Split Scheme.}} Controlling split operation is simpler because adding edges to existing faces for splitting does not change the original geometry. Nevertheless, the implementation of angular surveillance remains imperative. The reason is that performing a split operation may introduce lower-quality obtuse triangles, thereby degrading isotropic property. An instance is shown in Fig.~\ref{F1}. If determined solely by edge length criteria, the edge $\overline{p_1p_2}$ should be split. However, since angles $\angle\alpha$ and $\angle\beta$ in the original face are obtuse, performing the split would generate new triangles with even larger obtuse angles $\angle\alpha'$, $\angle\beta'$. These highly distorted triangular elements would adversely affect the convergence of isotropic remeshing. Therefore, it is necessary to evaluate whether any of its four adjacent angles ($\angle\alpha, \angle\beta, \angle p_3p_2p_1$, and $\angle p_4p_2p_3$) are obtuse before splitting.

\begin{figure}
\centering
\vspace{-4mm}
\includegraphics[width=0.9\textwidth]{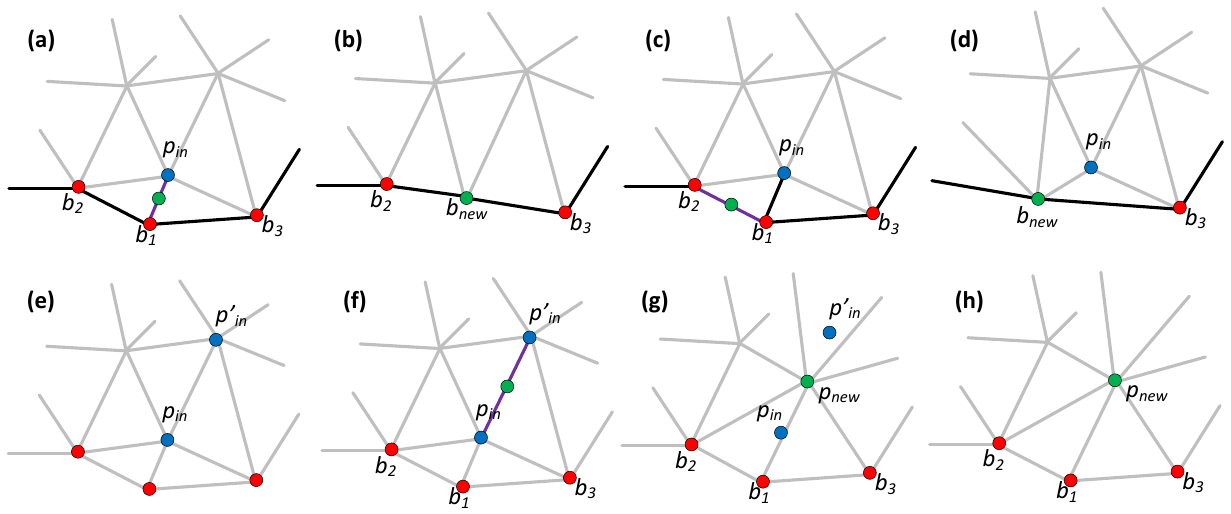}
\vspace{-2mm}
\caption{Inter-angle optimization for collapse scheme. (a)$\sim$(d) show the boundary condition; (e)$\sim$(h) show the degree judgment.} 
\vspace{-6mm}
\label{F2}
\end{figure}

\textbf{\textit{Collapse Scheme.}} In practice, the collapse operation induces the most significant geometric alterations, not only reducing the vertex count but also directly modifying vertex-based degrees, while causing substantial angular distortions. For collapse scheme, we employ two additional constraints: boundary condition and degree judgment. The boundary condition means that if one point of an edge belong to a boundary edge, then the collapse operation should be blocked, as shown in Fig.~\ref{F2}. Edge points $b_1$, $b_2$, and $b_3$ connect a boundary line labeled by black color. Once we collapse either $\overline{p_{in}b_1}$(Fig.~\ref{F2}a) or $\overline{b_1b_2}$(Fig.~\ref{F2}c), the geometry of the boundary line is changed, which breaks the geometric consistency and becomes severe with more iterations.

Another constraint is the degree judgment, which estimates the impact of the collapse operation on vertex degrees to determine whether to proceed. Without vertex-based degree analysis may lead to abnormal degree assignments for the new vertex, consequently increasing the difficulty of subsequent flip operations. From the perspective of inter-angle view, an excessively high vertex degree implies that the average inter-angle associated with that vertex will fall below the ideal threshold ($60^\circ$). It means the collapse operation may disrupt the isotropic property of the local region. An instance is also shown in Fig.~\ref{F2}. The degree values of $p_{in}$ and $p_{in}'$ are $deg(p_{in})=5$ and $deg(p_{in}')=6$ (average inter-angle is $65^\circ$, see in Fig.~\ref{F2}e). Once the collapse operation is implemented, the degree of new point $p_{new}$ change to $deg(p_{new})=7$ (average inter-angle is $50^\circ$, see in Fig.~\ref{F2}h). Evidently, performing collapse on $\overline{p_{in}p_{in}'}$ reduces the isotropic quality. Combined mentioned constraints, collapse scheme is completed.

\begin{figure}
\centering
\includegraphics[width=0.85\textwidth]{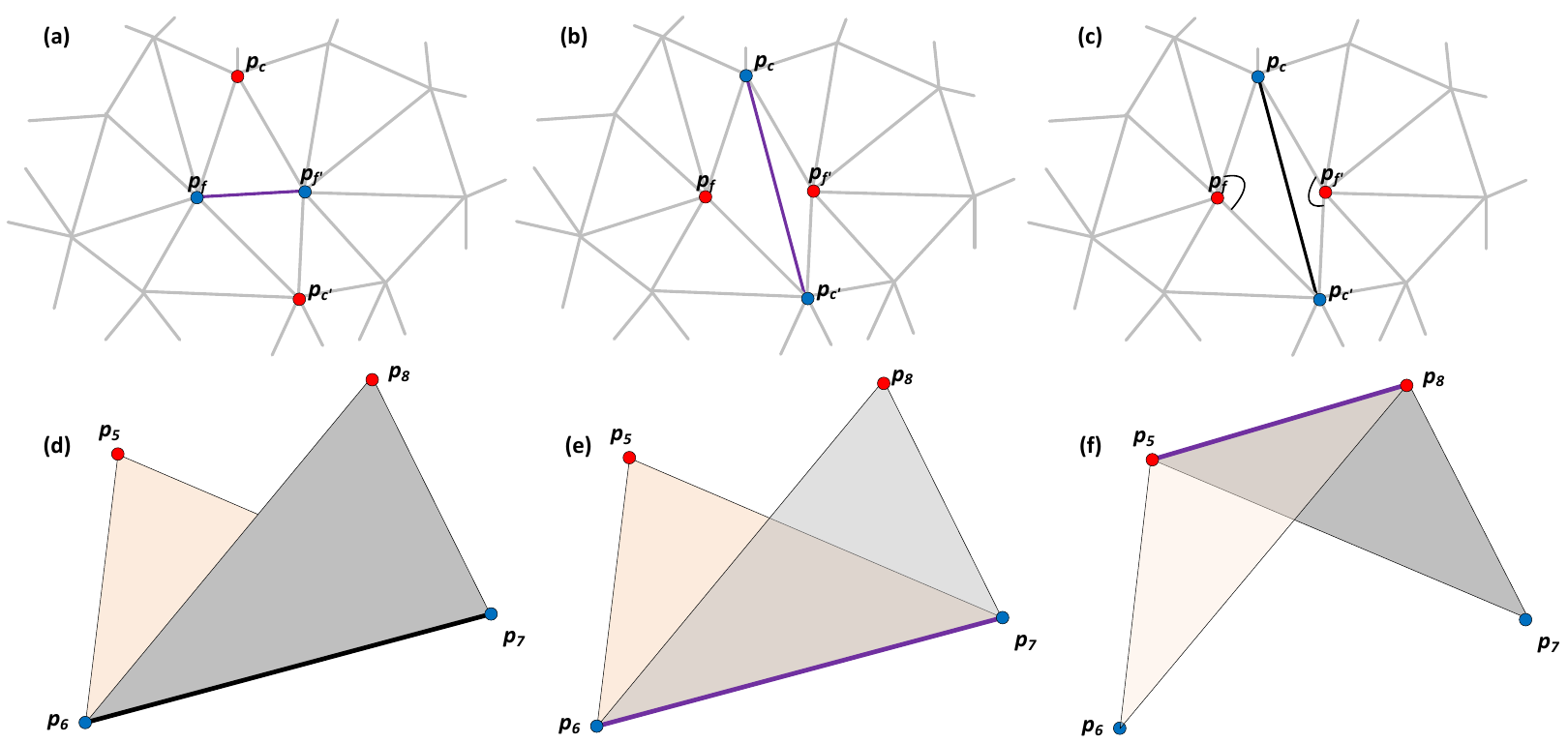}
\vspace{-2mm}
\caption{Inter-angle optimization for flip scheme. (a)$\sim$(c) show the inter-angle constraint; (e)$\sim$(f) show the shape control.} 
\vspace{-6mm}
\label{F3}
\end{figure}

\textbf{\textit{Flip Scheme.}} Beyond the primary objective of degree optimization, flip operation must account the inter-angle constraint, which is similar to the split scheme. An instance is shown in Fig~\ref{F3}. When determining whether to flip a given edge $\overline{p_fp_{f}'}$, we first compute the post-flip degree for each affected vertex. It is clear that the flip can optimize most vertexes, $deg(p_{c})=6, deg(p_{f})=6, deg(p_{f}')=6$, as shown in Fig.~\ref{F3}b. However, generated new obtuse angles ($\angle{p_cp_fp_{c}'}$ and $\angle{p_cp_{f}'p_{c}'}$ in Fig.~\ref{F3}) violate shape control requirements in the isotropic remeshing. Therefore, the flip operation should be blocked for such condition.

An additional consideration is the geometric alteration induced by the flip operation. In fact, unless all four vertices involved in the flip operation are coplanar, it inevitably modifies the local geometry, as shown in Fig.~\ref{F3}e and Fig.~\ref{F3}f. If edge $\overline{p_6p_7}$ lies precisely on a sharp feature boundary, performing the flip operation would introduce geometric artifacts along the boundary, thereby breaking the sharp feature. To mitigate the issue, we implement a face-based dihedral angle checking. Based on the instance in Fig.~\ref{F3}d, we compute the normal vectors of $\triangle p_5p_6p_7$ and $\triangle p_8p_7p_6$, and judge the vector-based intersection angle $\theta$. Once $\theta>\varepsilon$($\varepsilon=20^\circ$ by default), the flip operation is blocked for shape controlling. Combined split, collapse, and flip Schemes, the complete inter-angle optimization is established.

\subsection{Geometric Consistency Keeping}

The last operation of the four-steps remeshing is the tangent smoothing or vertex relocation, which adjusts vertex positions toward to their 1-ring neighborhood centers (gravity-weighted centroid~\cite{botsch2004remeshing}). The center can be formulated as
\begin{equation}
p_{i}'=\sum_{p_j\in N(p_i)}w_jp_j, 
\end{equation}
where $p_{i}'$ is the center for the region of point $p_{i}$, $p_j$ is 1-ring neighbor of $p_{i}$, $w_j$ is the weight of $p_j$, which is typically represented by the area or cotangent weights corresponding to $p_j$ to describe its influence on $p_i$. According to the vector $\overrightarrow{p_ip_{i}'}$(mapped on the tangent plane of $p_i$) with a "pulling back" function, $p_{i}$ can be updated
\begin{equation}
p_{i}'=p_i+\lambda(I-n_in_i^{T})\overrightarrow{p_ip_{i}'}, 
\end{equation}
where $n_i$ represent the normal vector of $p_i$, $\lambda$ is the control parameter for update step size ($\lambda=0.5$ by default). The aforementioned process is a common practice for vertex optimization while preserving geometric consistency. However, if the vertices of the original mesh are relatively sparse, the "pulling back" operation may fail to produce correct surface mapping due to ambiguity in neighborhood determination, resulting in local geometric distortion.

To address the issue, we propose a mesh up-sampling step to insert new points into different faces. Firstly, inserting the centroid of a triangular face. Then, based on the centroid, inserting the midpoints of the line segments between each vertex of the triangle and the centroid. Subsequently, inserting new centers between these newly created midpoints. Then, seven new points can be inserted for a single triangle. We adjust these new points‘ positions by MLS remapping~\cite{alexa2001point}. Finally, we achieve an up-sampled point set that takes more accurate geometric representation. Once we implement the tangent smoothing, the "pulling back" function is based on the neighborhood defined on the up-sampled point set. The local geometric distortion can be significantly suppressed. 

\section{Experiments}

In this part, We evaluate the performance of our remeshing scheme. The experimental machine equipped with Intel(R) i9-13900K 3.00 GHz, 128G RAM, GeForce RTX4090. The running system is Windows 11 with Visual Studio 2022 (64 bit). Firstly, we introduce the employed dataset and metrics. Then, we compare the performance between different methods. Finally, we show some applications based on our remeshing scheme and provide a discussion. 

\subsubsection{\textbf{Dataset\&Metrics.}} The experimental raw meshes are selected from SHREC~\cite{bronstein2010shrec}, including non-rigid transformations. For quantitative analysis, we report Hausdorff distance, mean distance, average and maximum inter-angle values to measure the geometric consistency and isotropic property.  

\begin{figure}[t]
\centering
\includegraphics[width=\textwidth]{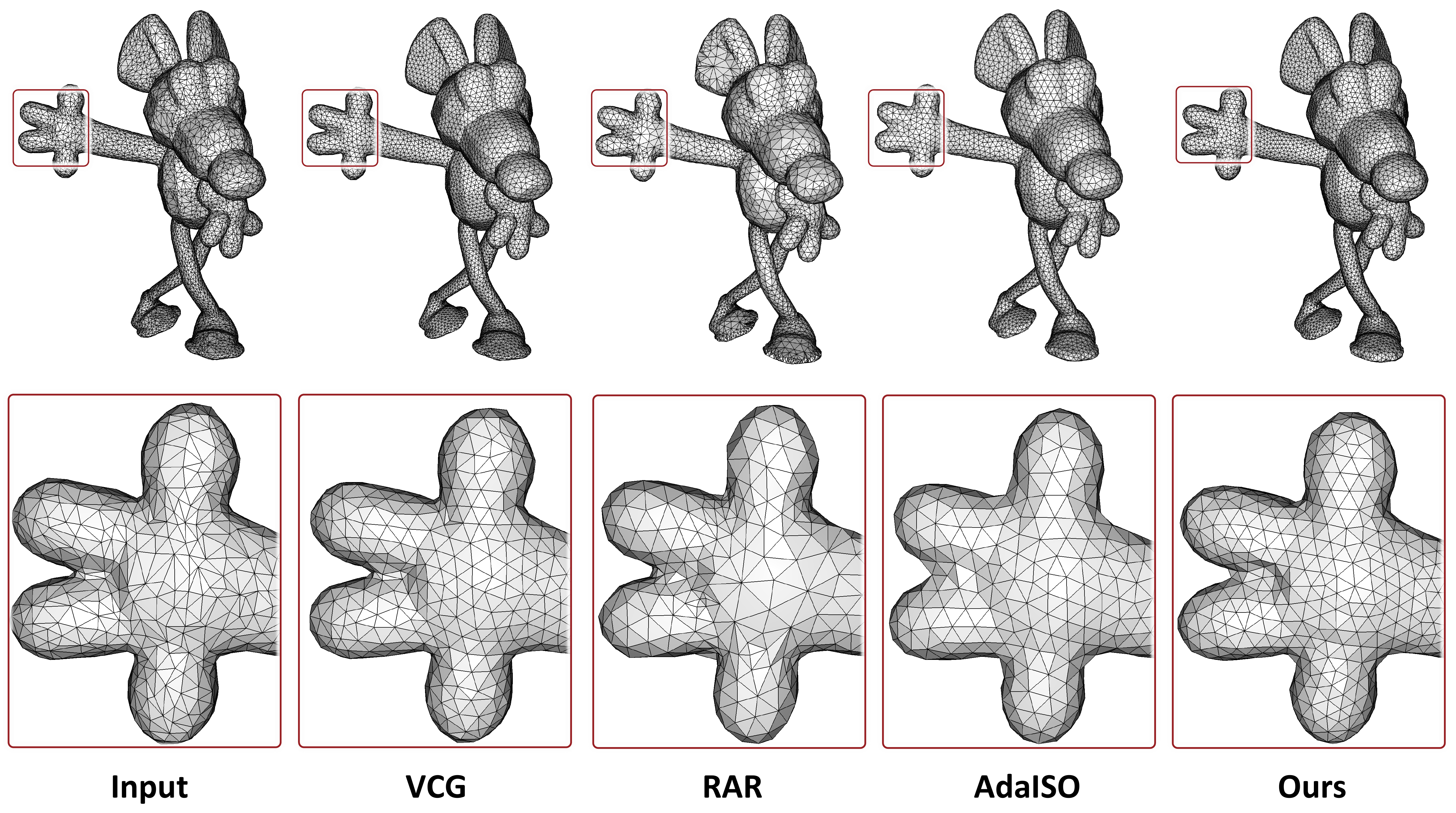}
\vspace{-6mm}
\caption{Some remeshing results by different methods.} 
\vspace{-2mm}
\label{F3_vis}
\end{figure}

\begin{figure}[t]
\centering
\includegraphics[width=\textwidth]{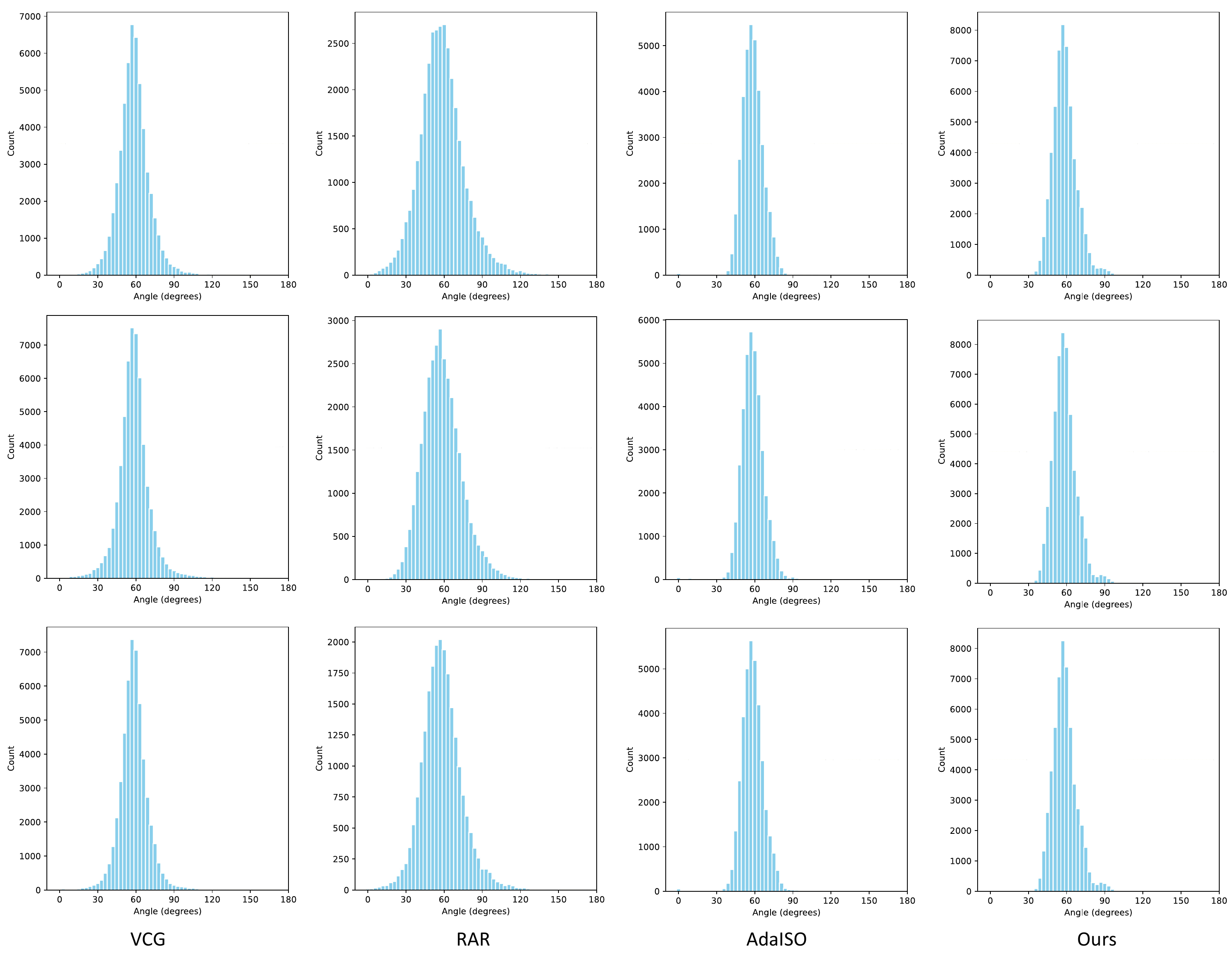}
\vspace{-8mm}
\caption{Inter-angle distribution histograms of re-meshed models (T0, T14, T28).} 
\vspace{-2mm}
\label{F4}
\end{figure}

\begin{table}[t]
\centering
\caption{Quantitative analysis of different remeshing methods. Hd: Hausdorff distance; Md: mean distance; $\theta_{max}$: maximum degree of inter-angel; $\theta_{avg}$: mean degree of inter-angel ($\pi/3-\sum\left|\theta_i-\pi/3\right|$, $\theta_i$ is an inter-angel).}
\label{T1}
\setlength{\tabcolsep}{6pt}    
\resizebox{\columnwidth}{!}{
\begin{tabular}{|l|cccc|cccc|cccc|cccc|}
\hline
\multirow{2}{*}{\textbf{Model}} & \multicolumn{4}{c|}{\textbf{VCG}~\cite{botsch2004remeshing}}                                                                               & \multicolumn{4}{c|}{\textbf{RAR}~\cite{dunyach2013adaptive}}                                                                               & \multicolumn{4}{c|}{\textbf{AdaISO}~\cite{lv2022adaptively}}                                                                                     & \multicolumn{4}{c|}{\textbf{Ours}}                                                                                                         \\ \cline{2-17} 
                                             & \multicolumn{1}{c|}{Hd}     & \multicolumn{1}{c|}{Md}              & \multicolumn{1}{c|}{$\theta_{max}$}  & $\theta_{avg}$ & \multicolumn{1}{c|}{Hd}              & \multicolumn{1}{c|}{Md}     & \multicolumn{1}{c|}{$\theta_{max}$}  & $\theta_{avg}$ & \multicolumn{1}{c|}{Hd}              & \multicolumn{1}{c|}{Md}     & \multicolumn{1}{c|}{$\theta_{max}$}  & $\theta_{avg}$          & \multicolumn{1}{c|}{Hd}              & \multicolumn{1}{c|}{Md}              & \multicolumn{1}{c|}{$\theta_{max}$}           & $\theta_{avg}$          \\ \hline
T0                                           & \multicolumn{1}{c|}{0.0159} & \multicolumn{1}{c|}{0.0016}          & \multicolumn{1}{c|}{151.5$^\circ$} & 50.9$^\circ$ & \multicolumn{1}{c|}{\textbf{0.0130}} & \multicolumn{1}{c|}{0.0016} & \multicolumn{1}{c|}{179.1$^\circ$} & 46.5$^\circ$ & \multicolumn{1}{c|}{0.0157}          & \multicolumn{1}{c|}{0.0017} & \multicolumn{1}{c|}{179.9$^\circ$} & 52.7$^\circ$          & \multicolumn{1}{c|}{0.0179}          & \multicolumn{1}{c|}{\textbf{0.0015}} & \multicolumn{1}{c|}{\textbf{118.7$^\circ$}} & \textbf{52.8$^\circ$} \\ \hline
T1008                                        & \multicolumn{1}{c|}{0.0127} & \multicolumn{1}{c|}{0.0017}          & \multicolumn{1}{c|}{167.2$^\circ$} & 50.5$^\circ$ & \multicolumn{1}{c|}{\textbf{0.0115}} & \multicolumn{1}{c|}{0.0018} & \multicolumn{1}{c|}{166.8$^\circ$} & 48.7$^\circ$ & \multicolumn{1}{c|}{0.0116}          & \multicolumn{1}{c|}{0.0017} & \multicolumn{1}{c|}{179.9$^\circ$} & \textbf{53.7$^\circ$} & \multicolumn{1}{c|}{0.0129}          & \multicolumn{1}{c|}{\textbf{0.0016}} & \multicolumn{1}{c|}{111.1$^\circ$}          & 52.9$^\circ$          \\ \hline
T1022                                        & \multicolumn{1}{c|}{0.0076} & \multicolumn{1}{c|}{0.0010}          & \multicolumn{1}{c|}{164.8$^\circ$} & 50.3$^\circ$ & \multicolumn{1}{c|}{——}              & \multicolumn{1}{c|}{——}     & \multicolumn{1}{c|}{——}    & ——   & \multicolumn{1}{c|}{\textbf{0.0070}} & \multicolumn{1}{c|}{0.0011} & \multicolumn{1}{c|}{179.9$^\circ$} & \textbf{52.6$^\circ$} & \multicolumn{1}{c|}{0.0072}          & \multicolumn{1}{c|}{\textbf{0.0009}} & \multicolumn{1}{c|}{\textbf{146.2$^\circ$}} & 52.5$^\circ$          \\ \hline
T1036                                        & \multicolumn{1}{c|}{0.0121} & \multicolumn{1}{c|}{0.0016}          & \multicolumn{1}{c|}{145.8$^\circ$} & 53.2$^\circ$ & \multicolumn{1}{c|}{0.0117}          & \multicolumn{1}{c|}{0.0016} & \multicolumn{1}{c|}{159.1$^\circ$} & 46.3$^\circ$ & \multicolumn{1}{c|}{\textbf{0.0111}} & \multicolumn{1}{c|}{0.0017} & \multicolumn{1}{c|}{179.9$^\circ$} & \textbf{53.5$^\circ$} & \multicolumn{1}{c|}{0.0137}          & \multicolumn{1}{c|}{\textbf{0.0016}} & \multicolumn{1}{c|}{\textbf{110.7$^\circ$}} & 53.1$^\circ$          \\ \hline
T1050                                        & \multicolumn{1}{c|}{0.0097} & \multicolumn{1}{c|}{0.0011}          & \multicolumn{1}{c|}{172.5$^\circ$} & 51.7$^\circ$ & \multicolumn{1}{c|}{\textbf{0.0089}} & \multicolumn{1}{c|}{0.0013} & \multicolumn{1}{c|}{154.4$^\circ$} & 49.3$^\circ$ & \multicolumn{1}{c|}{0.0090}          & \multicolumn{1}{c|}{0.0012} & \multicolumn{1}{c|}{179.9$^\circ$} & \textbf{53.4$^\circ$} & \multicolumn{1}{c|}{0.0101}          & \multicolumn{1}{c|}{\textbf{0.0011}} & \multicolumn{1}{c|}{\textbf{172.8$^\circ$}} & 53.1$^\circ$          \\ \hline
T1064                                        & \multicolumn{1}{c|}{0.0231} & \multicolumn{1}{c|}{0.0026}          & \multicolumn{1}{c|}{177.4$^\circ$} & 49.7$^\circ$ & \multicolumn{1}{c|}{0.0201}          & \multicolumn{1}{c|}{0.0028} & \multicolumn{1}{c|}{176.8$^\circ$} & 44.1$^\circ$ & \multicolumn{1}{c|}{\textbf{0.0176}} & \multicolumn{1}{c|}{0.0028} & \multicolumn{1}{c|}{180.0$^\circ$} & \textbf{53.1$^\circ$} & \multicolumn{1}{c|}{0.0202}          & \multicolumn{1}{c|}{\textbf{0.0026}} & \multicolumn{1}{c|}{\textbf{151.6$^\circ$}} & 52.7$^\circ$          \\ \hline
T1078                                        & \multicolumn{1}{c|}{0.0146} & \multicolumn{1}{c|}{0.0020}          & \multicolumn{1}{c|}{176.9$^\circ$} & 47.7$^\circ$ & \multicolumn{1}{c|}{0.0156}          & \multicolumn{1}{c|}{0.0020} & \multicolumn{1}{c|}{174.8$^\circ$} & 48.1$^\circ$ & \multicolumn{1}{c|}{——}              & \multicolumn{1}{c|}{——}     & \multicolumn{1}{c|}{——}    & ——            & \multicolumn{1}{c|}{0.0159}          & \multicolumn{1}{c|}{\textbf{0.0020}} & \multicolumn{1}{c|}{\textbf{167.7$^\circ$}} & \textbf{52.6$^\circ$} \\ \hline
T1092                                        & \multicolumn{1}{c|}{0.0165} & \multicolumn{1}{c|}{0.0017}          & \multicolumn{1}{c|}{169.9$^\circ$} & 52.9$^\circ$ & \multicolumn{1}{c|}{\textbf{0.0162}} & \multicolumn{1}{c|}{0.0018} & \multicolumn{1}{c|}{177.1$^\circ$} & 46.5$^\circ$ & \multicolumn{1}{c|}{0.0163}          & \multicolumn{1}{c|}{0.0019} & \multicolumn{1}{c|}{180.0$^\circ$} & 53.2$^\circ$          & \multicolumn{1}{c|}{0.0180}          & \multicolumn{1}{c|}{\textbf{0.0017}} & \multicolumn{1}{c|}{\textbf{106.0$^\circ$}} & \textbf{53.3$^\circ$} \\ \hline
T1106                                        & \multicolumn{1}{c|}{0.0105} & \multicolumn{1}{c|}{0.0014}          & \multicolumn{1}{c|}{178.2$^\circ$} & 50.7$^\circ$ & \multicolumn{1}{c|}{0.0110}          & \multicolumn{1}{c|}{0.0016} & \multicolumn{1}{c|}{175.8$^\circ$} & 47.6$^\circ$ & \multicolumn{1}{c|}{0.0107}          & \multicolumn{1}{c|}{0.0015} & \multicolumn{1}{c|}{179.9$^\circ$} & \textbf{53.4$^\circ$} & \multicolumn{1}{c|}{\textbf{0.0096}} & \multicolumn{1}{c|}{\textbf{0.0014}} & \multicolumn{1}{c|}{\textbf{119.6$^\circ$}} & 53.1$^\circ$          \\ \hline
T112                                         & \multicolumn{1}{c|}{0.0244} & \multicolumn{1}{c|}{\textbf{0.0031}} & \multicolumn{1}{c|}{177.1$^\circ$} & 46.9$^\circ$ & \multicolumn{1}{c|}{——}              & \multicolumn{1}{c|}{——}     & \multicolumn{1}{c|}{——}    & ——   & \multicolumn{1}{c|}{——}              & \multicolumn{1}{c|}{——}     & \multicolumn{1}{c|}{——}    & ——            & \multicolumn{1}{c|}{\textbf{0.0227}} & \multicolumn{1}{c|}{0.0034}          & \multicolumn{1}{c|}{\textbf{145.1$^\circ$}} & \textbf{52.4$^\circ$} \\ \hline
T1120                                        & \multicolumn{1}{c|}{0.0155} & \multicolumn{1}{c|}{0.0023}          & \multicolumn{1}{c|}{175.7$^\circ$} & 49.3$^\circ$ & \multicolumn{1}{c|}{0.0177}          & \multicolumn{1}{c|}{0.0023} & \multicolumn{1}{c|}{174.4$^\circ$} & 46.3$^\circ$ & \multicolumn{1}{c|}{\textbf{0.0138}} & \multicolumn{1}{c|}{0.0024} & \multicolumn{1}{c|}{180.0$^\circ$} & 53.1$^\circ$          & \multicolumn{1}{c|}{0.0157}          & \multicolumn{1}{c|}{\textbf{0.0023}} & \multicolumn{1}{c|}{\textbf{137.1$^\circ$}} & \textbf{53.2$^\circ$} \\ \hline
T1134                                        & \multicolumn{1}{c|}{0.0086} & \multicolumn{1}{c|}{0.0012}          & \multicolumn{1}{c|}{175.3$^\circ$} & 49.6$^\circ$ & \multicolumn{1}{c|}{\textbf{0.0076}} & \multicolumn{1}{c|}{0.0013} & \multicolumn{1}{c|}{173.7$^\circ$} & 48.5$^\circ$ & \multicolumn{1}{c|}{——}              & \multicolumn{1}{c|}{——}     & \multicolumn{1}{c|}{——}    & ——            & \multicolumn{1}{c|}{0.0080}          & \multicolumn{1}{c|}{\textbf{0.0011}} & \multicolumn{1}{c|}{\textbf{138.6$^\circ$}} & \textbf{53.0$^\circ$} \\ \hline
T1155                                        & \multicolumn{1}{c|}{0.0165} & \multicolumn{1}{c|}{0.0020}          & \multicolumn{1}{c|}{173.1$^\circ$} & 51.7$^\circ$ & \multicolumn{1}{c|}{\textbf{0.0109}} & \multicolumn{1}{c|}{0.0021} & \multicolumn{1}{c|}{176.0$^\circ$} & 48.4$^\circ$ & \multicolumn{1}{c|}{0.0147}          & \multicolumn{1}{c|}{0.0021} & \multicolumn{1}{c|}{79.9$^\circ$}  & \textbf{53.4$^\circ$} & \multicolumn{1}{c|}{0.0177}          & \multicolumn{1}{c|}{\textbf{0.0020}} & \multicolumn{1}{c|}{\textbf{114.9$^\circ$}} & 53.0$^\circ$          \\ \hline
Avg                                          & \multicolumn{1}{c|}{0.0120} & \multicolumn{1}{c|}{0.0017}          & \multicolumn{1}{c|}{175.6$^\circ$} & 50.4$^\circ$ & \multicolumn{1}{c|}{0.0107}          & \multicolumn{1}{c|}{0.0017} & \multicolumn{1}{c|}{169.6$^\circ$} & 48.6$^\circ$ & \multicolumn{1}{c|}{\textbf{0.0105}} & \multicolumn{1}{c|}{0.0017} & \multicolumn{1}{c|}{179.7$^\circ$} & 52.9$^\circ$          & \multicolumn{1}{c|}{0.0116}          & \multicolumn{1}{c|}{\textbf{0.0016}} & \multicolumn{1}{c|}{\textbf{113.8$^\circ$}} & \textbf{53.0$^\circ$} \\ \hline
\end{tabular}}
\vspace{-6mm}
\end{table}

\begin{table}[h]
\centering
\caption{Time cost report for different methods.}
\label{T2}
\vspace{-2mm}
\setlength{\tabcolsep}{6pt} 
\begin{tabular}{|l|c|c|c|c|c|}
\hline
    & \textbf{VCG~\cite{botsch2004remeshing}} & \textbf{RAR~\cite{dunyach2013adaptive}} & \textbf{FCVT~\cite{du2018field}}& \textbf{AdaISO~\cite{lv2022adaptively}} & \textbf{Ours} \\ \hline
T0  & 19.86s           & 9.08s           & 400.25s   & 31.64s           & \textbf{8.93s }           \\ \hline
T14 & 21.26s           & \textbf{2.95s}   & 381.85s        & 39.72s              & 11.26s            \\ \hline
T28 & 23.56s           & \textbf{1.58s}    & 387.19s       & 43.03s              & 13.26s            \\ \hline
Avg & 20.63s           & —— & 391.21s            & 38.93s              & \textbf{10.89s}            \\ \hline
\end{tabular}
\end{table}

\subsubsection{\textbf{Comparisons.}} We compare different methods to show the performance of our remeshing scheme, including VCG-based isotropic remeshing (same to MeshLab version~\cite{botsch2004remeshing,Cignoni2008}), RAR remeshing~\cite{dunyach2013adaptive}, and adaptive isotropic remeshing~\cite{lv2022adaptively}. Fig.~\ref{F3_vis} shows some remeshing results by different methods. Our solution takes better balance between geometric consistency and isotropic property. Table~\ref{T1} reports quantitative results based on SHREC models. Our method demonstrates superior shape control for large obtuse triangles ($\theta_{max}$ is smaller). 

Benefiting from the implementation of geometric consistency keeping, our approach achieves better geometric consistency (Md value is lower). Fig.~\ref{F4} shows some  histograms of inter-angle distribution. AdaISO and our solution take better distributions (concentrated around 60$^\circ$). Table~\ref{T2} reports computational efficiency for different methods. Our solution is faster than VCG-based isotropic remeshing and AdaISO solution, even they share similar four-steps strategy. RAR solution achieves faster runtime speeds for certain models, but fails to converge for some ones based on a larger test dataset. In addition, we employ a CVT-based solution (FCVT)~\cite{du2018field} to be a reference in Table~\ref{T2}. According to statistical data, FCVT requires 30 times longer to achieve results comparable to our method. Significantly, our method holds advantages in terms of efficiency and stability.

\subsubsection{\textbf{Applications.}} 
\begin{figure}[t]
\centering
\includegraphics[width=\textwidth]{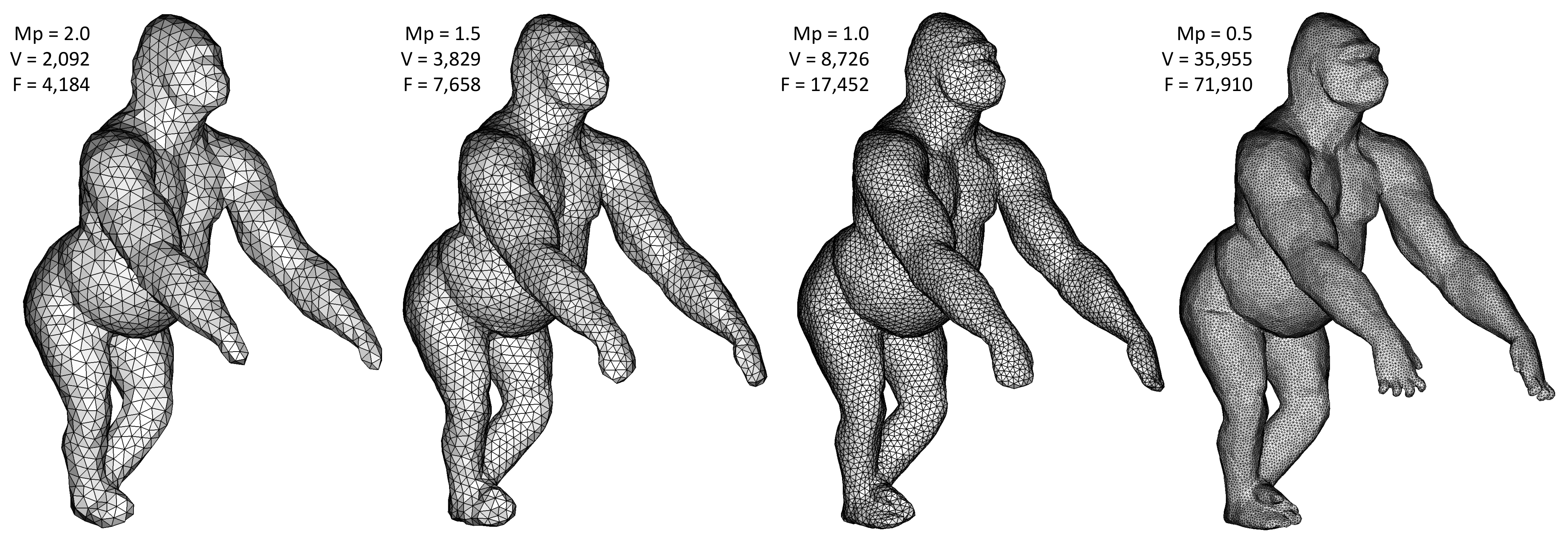}
\vspace{-6mm}
\caption{Remeshing results by our method with different multi-parameters. Mp: multi-parameter value; V: vertex number; F: face number.} 
\vspace{-2mm}
\label{F6_multi}
\end{figure}

\begin{figure}[t]
\centering
\includegraphics[width=0.8\textwidth]{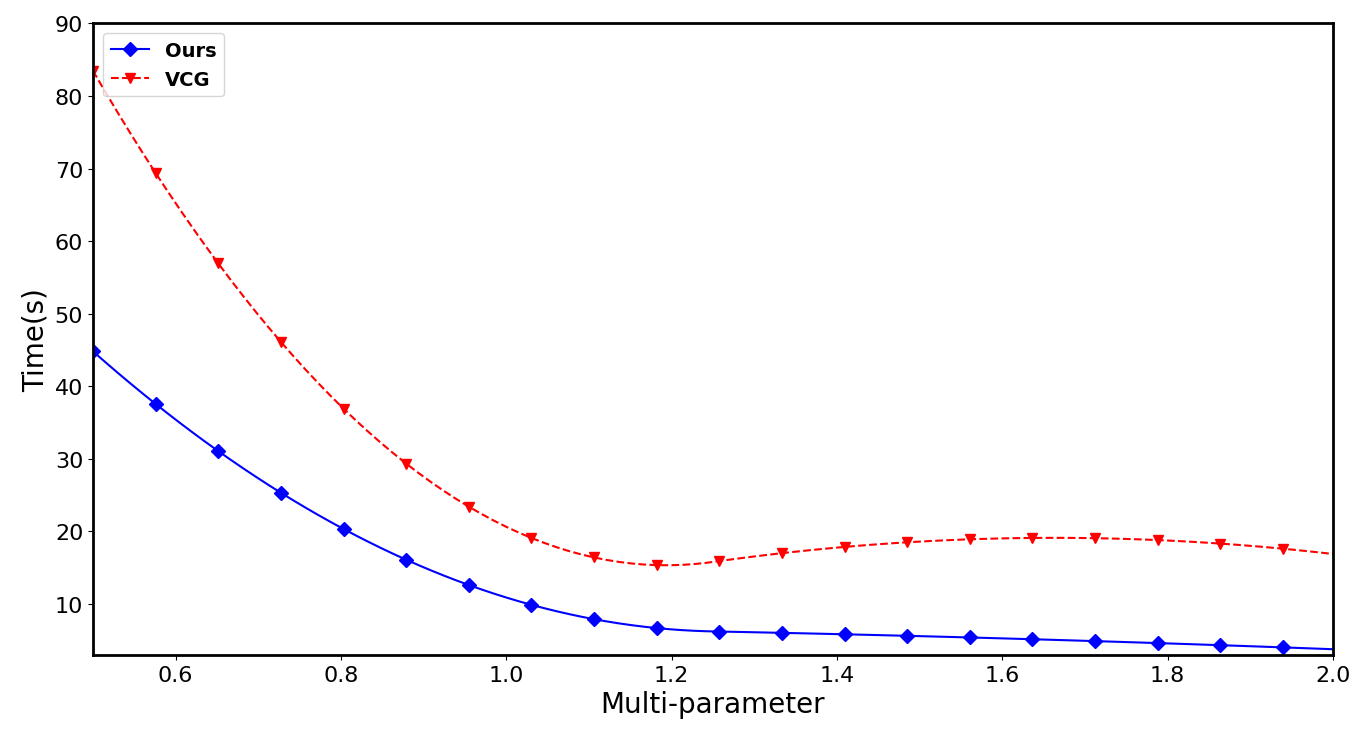}
\vspace{-4mm}
\caption{Time cost curves for different methods with related multi-parameters.} 
\vspace{-2mm}
\label{F7_time}
\end{figure}

A classical application of remeshing is the multi-resolution editing. By modifying the target edge length, we can either simplify or refine the input mesh. We assign different multi-parameters and  multiply them by the average edge length to obtain new target edge lengths, thereby generating remeshing results with varying resolutions. Fig.~\ref{F6_multi} shows some remeshing results by our method with different multi-parameters. The mesh resolution can be adjusted while preserving the isotropic property. We also report the time cost of remeshing with continuous multi-parameter variations, and add VCG-based remeshing as the reference. Fig.~\ref{F7_time} shows related time cost curves. Our method demonstrates superior time efficiency in both mesh simplification and refinement.

\begin{figure}[t]
\centering
\includegraphics[width=\textwidth]{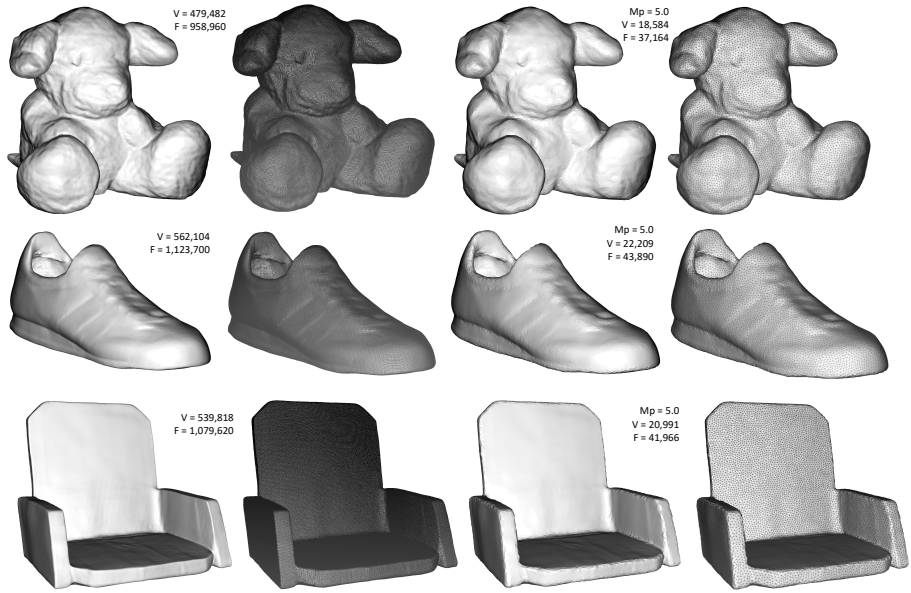}
\vspace{-4mm}
\caption{Remeshing results based on 3D mesh generation framework. Left: generated meshes; right: remeshing results.} 
\vspace{-2mm}
\label{F8_AIGC}
\end{figure}

Another important application is assisting 3D-AIGC tasks. Currently, mainstream 3D mesh generation methods primarily rely on implicit reconstruction, driven by semantic and image features to produce accurate mesh outputs. However, they suffer from poor isotropic quality and excessive mesh volume. Our method can effectively enhance the isotropic quality of meshes while simultaneously simplifying them. Based on the generated results by MeshFormer~\cite{liu2024meshformer}, our remeshing is used to improve the mesh quality. Fig.~\ref{F8_AIGC} shows some instances. It can be clearly observed that both the mesh volume and isotropic quality are optimized.

\subsubsection{\textbf{Discussion.}} Compared to the VCG-based remeshing solution, our method utilizes inter-angle control to avoid cross-interference between each basic operation, thereby improving convergence efficiency. Tables~\ref{T1} and \ref{T2} demonstrate that our approach outperforms VCG-based version in both mesh optimization quality and computational efficiency. Compared to the CVT-based approach, our method inherits the advantages of isotropic remeshing while demonstrating significant superiority in convergence speed. The core reason is that our basic operations enable more direct updates to triangle shapes. In contrast, the CVT-based method relies on Voronoi cell computation and centroidal optimization for mesh editing, which inherently suffers from efficiency limitation.

\section{Conclusions}

In this paper, we propose a new isotropic remeshing method with inter-angle
optimization. Angle-based constraints are incorporated into the basic operations (split, collapse, and flip) to suppress the generation of obtuse triangles. Compared to the original four-steps isotropic remeshing, our method effectively reduces repetitive edge editing and improves convergence efficiency. Benefited from a mesh-based up-sampling scheme, the quality of isotropic remeshing can be significantly improved. Experimental reports demonstrate that our solution exhibits comprehensive advantages in isotropic property optimization, geometric consistency keeping, and computational efficiency.

%
%
\bibliographystyle{splncs04}
%
\bibliography{ref}

\end{document}